# Stabilization of active matter by flow-vortex lattices and defect ordering

Amin Doostmohammadi[1], Michael F. Adamer[1], Sumesh P. Thampi[1] & Julia M. Yeomans[1]

Active systems, from bacterial suspensions to cellular monolayers, are continuously driven out of equilibrium by local injection of energy from their constituent elements and exhibit turbulent-like and chaotic patterns. Here we demonstrate both theoretically and through numerical simulations, that the crossover between wet active systems, whose behaviour is dominated by hydrodynamics, and dry active matter where any flow is screened, can be achieved by using friction as a control parameter. Moreover, we discover unexpected vortex ordering at this wet–dry crossover. We show that the self organization of vortices into lattices is accompanied by the spatial ordering of topological defects leading to active crystal-like structures. The emergence of vortex lattices, which leads to the positional ordering of topological defects, suggests potential applications in the design and control of active materials.

[1] The Rudolf Peierls Centre for Theoretical Physics, 1 Keble Road, Oxford OX1 3NP, UK. Correspondence and requests for materials should be addressed to A.D. (email: amin.doostmohammadi@physics.ox.ac.uk).

Whether it is biological matter such as cytoskeletal networks, cellular colonies and suspensions of bacteria or synthetic systems such as Janus catalysts and vibrating granular rods, continuous injection of energy from the constituent elements leads to exotic behaviour such as collective motion[1–3], pattern formation[4–7], topological defects[8–10] and active turbulence[11–13]. Active systems where hydrodynamic interactions are key, such as suspensions of swimming bacteria, are referred to as wet. The energy input to such systems is dissipated by the viscosity of the fluid. However, in many active materials there are alternative forms of dissipation, such as wall friction between the active particles and a substrate[14]. In the limit where the frictional damping dominates, it screens out any velocity fields that are generated by the activity. For instance, the Vicsek model for migrating animal herds[15], and assemblies of vibrated granular rods[16,17] lack long-range hydrodynamic interactions and are classified as dry active materials. In general, a given active material will fall between these limits with the relative magnitudes of frictional and dissipative damping controlling its position on the wet–dry spectrum. Examples of experimental two-dimensional active nematics where friction could be varied to control the crossover between the wet and dry limits include cells moving and dividing on substrates[18,19], microswimmers confined between parallel walls[11] and suspensions of microtubules and molecular motors moving at an interface between two fluids[6].

Here we unify two different classes of active matter by using friction as a control parameter to interpolate between wet and dry active matter, and at the wet–dry crossover we discover an unexpected regime, where otherwise disordered vortices self-organize into lattices interleaved with ordered arrays of topological defects. The vortex lattices and their corresponding network of ordered defects arise from the competition between friction and viscous dissipation and stablize the active system. Our results contribute to understanding the physics of matter operating out of equilibrium, with its potential in the design of active micro- and nano-machines.

## Results

**Wet–dry crossover.** The behaviour of dry and wet active nematics is summarized pictorially at the right- and left-hand sides of Fig. 1. In the absence of momentum conservation, dry active matter is commonly described in terms of the concentration of active components and their orientational order[20–22]. The existence of any curvature in the orientational order induces an instability that leads to the formation of bands of concentration[21,22]. The concentration bands are, however, always unstable and eventually break into filaments that, in turn, coalesce and form new bands. The process is repeated and a chaotic regime is established[23].

Introducing hydrodynamics leads to extra complexity due to the coupling of the density and orientational order of active particles to the fluid flow, and distinct dynamical features are manifest in the models of wet active nematics. In particular, there is a hydrodynamic instability of nematic regions that leads to the formation of walls, lines of high distortions in the director field, where elastic energy is stored. The walls are continuously broken up by the creation and annihilation of topological defects, a process that gives rise once again to unstable nematic regions. The formation and removal of the walls is maintained by the active forcing, and a state termed active turbulence is established, characterized by a chaotic velocity field with regions of high vorticity[9,24].

**Theory.** We build on the nematohydrodynamic equations of liquid crystals to describe a natural route from wet to dry active nematics. To this end, the continuum description of the dynamics of passive liquid crystals is modified to account for the active stresses generated by constituent elements[25]. This continuum approach, which allows for coarse graining over the microscopic details, has proven successful in reproducing several experimental observations including the flow behaviour and defect dynamics observed in experiments on microtuble bundles driven by motor proteins[6,9,24], the spatial organization of bacterial cultures in confined environments[26], tissue growth[27,28] and fluidization[29], and the flow fields associated with cell division in cellular monolayers[30].

The variables needed to describe the hydrodynamics of a wet active nematic are: $\phi$ the relative concentration of active and passive particles, $\rho$ the total density, $\mathbf{u}$ the velocity vector, and $\mathbf{Q} = \frac{3q}{2}(\mathbf{nn} - \mathbf{I}/3)$, which is the nematic tensor, with $q$ the magnitude of the orientational order, $\mathbf{n}$ the director, and $\mathbf{I}$ the identity tensor. The four coupled continuum equations describing the time evolution of these quantities are[31]

$$\partial_t \phi + \partial_i(u_i \phi) = \Gamma_\phi \nabla^2 \mu, \quad (1)$$

$$(\partial_t + u_k \partial_k) Q_{ij} - S_{ij} = \Gamma_Q H_{ij}, \quad (2)$$

$$\partial_t \rho + \partial_i(\rho u_i) = 0, \quad (3)$$

$$\rho(\partial_t + u_k \partial_k) u_i = \partial_j \Pi_{ij} - \gamma u_i, \quad (4)$$

where the mobility and the rotational diffusivity are denoted by $\Gamma_\phi$ and $\Gamma_Q$, respectively. The co-rotation term,

$$S_{ij} = [\lambda E_{ik} + \Omega_{ik}]\left(Q_{kj} + \frac{\delta_{kj}}{3}\right) + \left(Q_{ik} + \frac{\delta_{ik}}{3}\right)[\lambda E_{kj} - \Omega_{kj}]$$
$$- 2\lambda\left(Q_{ij} + \frac{\delta_{ij}}{3}\right)(Q_{kl}\partial_k u_l), \quad (5)$$

accounts for the response of the orientational order to the flow gradients characterized by the strain rate tensor $E_{ij} = (\partial_i u_j + \partial_j u_i)/2$ and vorticity tensor $\Omega_{ij} = (\partial_j u_i - \partial_i u_j)/2$. The alignment parameter $\lambda$, which describes the responses of particles to the strain and vorticity takes different values for different particle shapes[32] with $\lambda > 0$, $\lambda < 0$ and $\lambda = 0$ for rod-like, disk-like and spherical particles, respectively. The chemical potential $\mu = \delta\mathcal{F}/\delta\phi$ and molecular field $H_{ij} = -\delta\mathcal{F}/\delta Q_{ij} + (\delta_{ij}/3)\mathrm{Tr}(\delta\mathcal{F}/\delta Q_{kl})$ represent the relaxation of the concentration and the orientational order to equilibrium, where

$$\mathcal{F} = \frac{a}{2}\phi^2 + \frac{A}{2}\phi\mathbf{Q}^2 + \frac{C}{4}\mathbf{Q}^4$$
$$+ \frac{K_\phi}{2}(\nabla\phi)^2 + \frac{K_Q}{2}(\nabla\mathbf{Q})^2 + K_c \mathbf{Q} : \nabla\nabla\phi, \quad (6)$$

is the free energy functional and $a$, $A$, $C$, $K_\phi$, $K_Q$, $K_c$ are material constants.

The terms on the left-hand side of equation (4) are the usual terms in the Navier–Stokes equation describing the inertia of the fluid. These are small at low Reynolds number, the limit relevant to active particles on the micron-scale or smaller. Contributions to the stress tensor, $\Pi_{ij}$, are the viscous dissipation $\Pi_{ij}^{\mathrm{viscous}} = 2\eta E_{ij}$, where $\eta$ is viscosity, the elastic stresses that generate backflow and encode mechanical forces on the fluid due to the relaxational motion of the active entities

$$\Pi_{ij}^{\mathrm{elastic}} = -P\delta_{ij} + (\mathcal{F} - \mu\phi)\delta_{ij}$$
$$+ 2\lambda\left(Q_{ij} + \frac{\delta_{ij}}{3}\right)(Q_{kl} H_{lk})$$
$$- \lambda H_{ik}\left(Q_{kj} + \frac{\delta_{kj}}{3}\right) - \lambda\left(Q_{ik} + \frac{\delta_{ik}}{3}\right)H_{kj} \quad (7)$$
$$- \partial_i Q_{kl}\frac{\delta\mathcal{F}}{\delta\partial_j Q_{lk}} - \partial_i\phi\frac{\delta\mathcal{F}}{\delta\partial_j\phi} + Q_{ik}H_{kj} - H_{ik}Q_{kj},$$

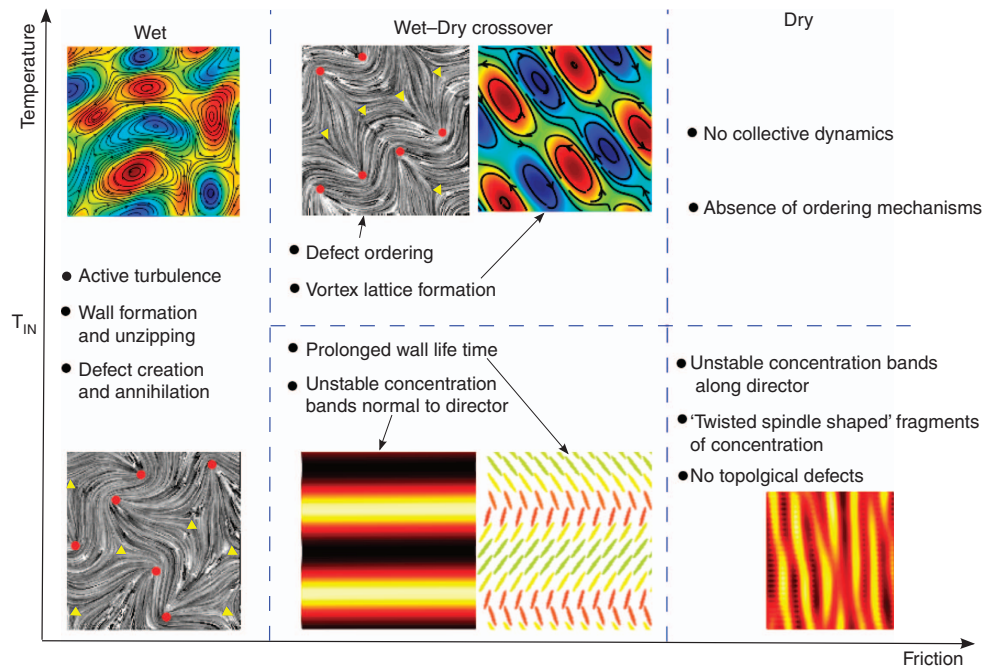

**Figure 1 | Dynamical behaviours of active nematics in the temperature-friction phase space.** $T_{IN}$ denotes the isotropic-nematic transition temperature: above $T_{IN}$ at intermediate frictions we find a novel vortex lattice that entangles an ordered defect array. The blue-red colourmaps represent vorticity fields superimposed by streamlines (black solid lines). The concentration fields are depicted by red-yellow colourmaps. In the mid-lower section, the director field is illustrated by ellipsoids coloured by their orientations. Grey panels show director fields superimposed by topological defects (red circles and yellow triangles correspond to $+1/2$ and $-1/2$ defects).

where $P$ denotes the pressure, and, of primary relevance to our current argument, the active stress $\Pi_{ij}^{active} = -\zeta Q_{ij}$, which is proportional to the orientational order parameter and distinguishes active from passive nematics[25]. The strength of the activity is set by $\zeta$. The final term in equation (4) accounts for the friction, for example, with a substrate, with $\gamma$, the associated friction coefficient. Details of this model in the context of passive and active liquid crystals can be found in refs 31,33–39.

In a wet active nematic, the momentum propagation is more effectively suppressed as the friction coefficient is increased. In the limit of sufficiently high friction, the viscous and elastic stresses become small and the friction forces balance with the forces generated by the active term. In this regime the velocity can be approximated in terms of the nematic tensor as

$$u_i \approx -\frac{\zeta}{\gamma}\partial_j Q_{ij}. \qquad (8)$$

In writing equation (8), we are neglecting the fluid inertia, which is dominated by the viscosity for bacterial suspensions and cytoskeletal filaments. We also neglect the elastic terms in the stress tensor that generate backflow. Non-dimensionalizing the momentum equation (see Methods section for more details) shows that the viscous stress dominates the backflow by a factor controlled by the Ericksen number $Er = \eta UL/K$, where $U$, $L$ are a characteristic velocity and length. $Er$ is $\sim 10^2$ in our simulations.

We chose parameters that are in a range that has been previously used to reproduce experimental results measuring the velocity correlations of microtuble bundles driven by molecular motors[6,24] and the flow fields of dividing Madine Darby Canine Kidney cells[30]. Moreover, the scaling suggested by equation (8), that velocity increases proportionately with activity, has been observed in both simulations[24] and experiments[6].

Considering equation (8) as an equality and substituting in the equations (1) and (2) for the concentration and orientational order evolution, we reproduce the standard equations for dry active matter,

$$\partial_t \phi = \Gamma_\phi \nabla^2 \mu + \frac{\zeta\phi}{\gamma}\partial_i\partial_j Q_{ij} + \ldots, \qquad (9)$$

$$\partial_t Q_{ij} = \Gamma_Q H_{ij} + \ldots \qquad (10)$$

The second term on the right-hand side of equation (9) is identical to that commonly introduced to represent the current due to curvature in the director field, which drives ordering in dry active nematics. Thus, we demonstrate that the equations of dry active nematics can be generically reproduced from the nematohydrodynamic equations, showing that the dry limit arises naturally when friction dominates viscosity in wet active materials. We use $+ \ldots$ to denote that flow-driven terms that are non-linear in $\phi$ and $\mathbf{Q}$ also appear in equation (9) and (10). These are listed explicitly in the Methods section and their appearance can be expected from symmetry arguments. In addition, we have also used a kinetic approach[21] to show that our analytical derivation of the dry active nematic equations from equations of wet active nematics, matches the microscopic derivation at the kinetic equation level. We show that even higher order terms predicted by our derivation can be expected from the expansion of the kinetic equations to higher order (Methods section).

**Simulation.** To demonstrate the wet to dry crossover, we present numerical solutions of the active nematic equations. We first illustrate the effect of increasing the friction on the evolution of the concentration and order parameter in a strongly ordered nematic (Fig. 2). Figure 2a shows the time evolution of the concentration field (top row) and director field (bottom row) for an active nematic with an intermediate friction coefficient. As for zero-friction the nematic state is hydrodynamically unstable and walls form in the director field. However, both wall creation, and subsequent destruction by topological defects, are significantly slowed by the friction, which leads to a reduction in the root-mean-square (RMS) velocity by a factor $\sim 10$ (Fig. 2c). Surprisingly, the slower dynamics is accompanied by an increase in the number of topological defects (Fig. 2d), a consequence of the larger number of walls at higher friction. Similar trends have been recently observed in active nematics without any concentration variation[40]. However, as described earlier, concentration variation is necessary to approach the dry limit where activity manifests in concentration phase separation.

Linked to the wall formation is the emergence of coincident concentration bands, with a higher number of active particles at the walls, where the magnitude of the nematic ordering is reduced. The concentration ordering is driven by an advective flux of active particles towards the walls, down the gradient in **Q**. In the steady state this is balanced by diffusion of the active particles from high to low concentrations.

Figure 2c,d show that the active system behaves very differently for higher values of the friction. The RMS velocity is very low and topological defects are not formed. We note that although, topological defects do not appear in the dry limit, such defects have been observed in shaken rod experiments and in a simulation of active rods. As pointed out in ref. 41 this occurs, in their model at least, because the collisions include a relative rotation of the colliding particles as well as nematic alignment. Figure 2b indicates that concentration bands still appear in the dry limit but, in the absence of an advective flux, the way in which they are formed must be very different. The relevant mechanism is an instability driven by curvature in the nematic order, described by the final term in equation (9). The coupling between the concentration and the nematic order is established by the molecular potential resulting in a strong (weak) ordering at high

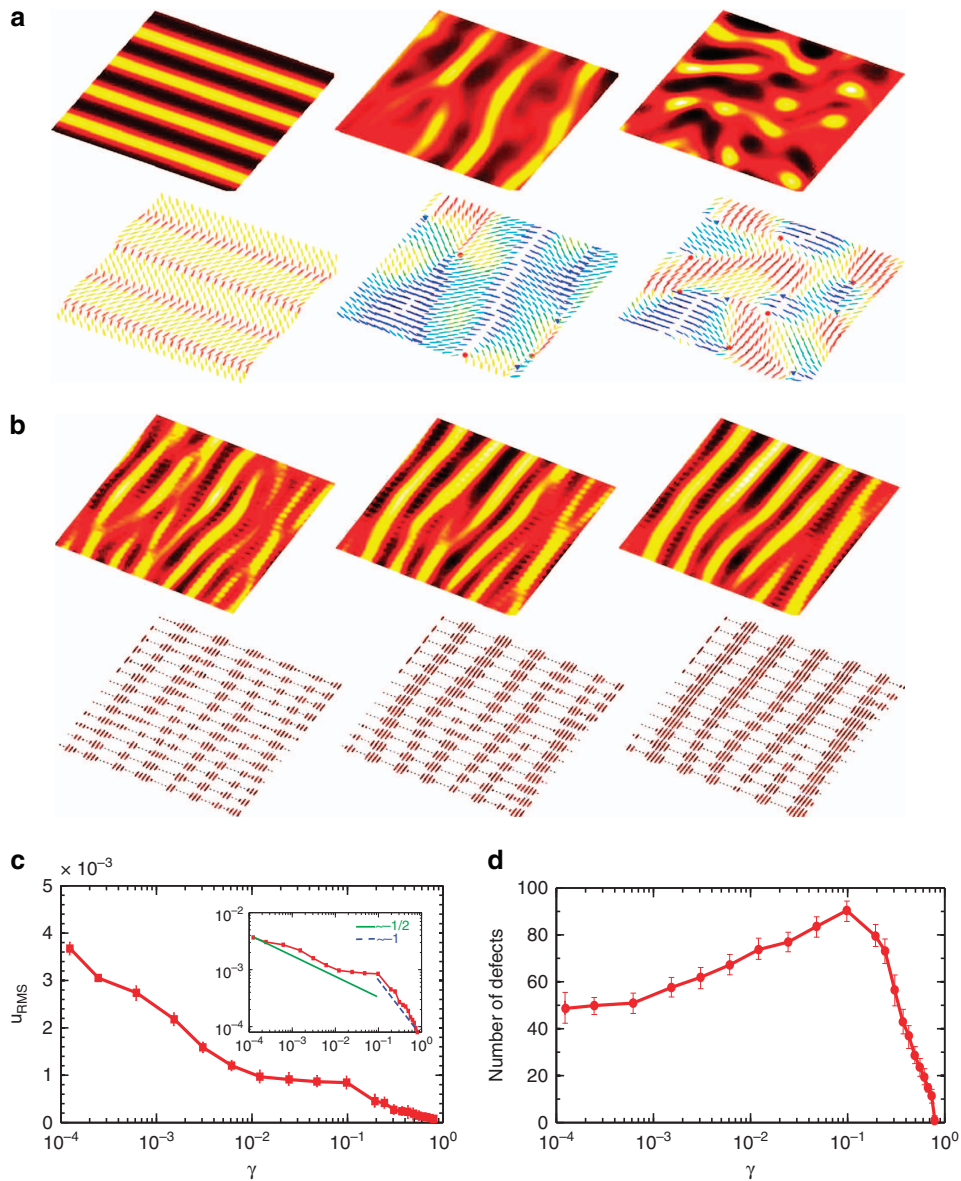

**Figure 2 | Increasing friction drives the crossover from wet to dry active nematics.** (**a,b**) Temporal evolution of the concentration field and nematic director field for (**a**) $\gamma = 0.08$ and (**b**) $\gamma = 0.8$ with $t = 10^5$, $5 \times 10^5$, $8 \times 10^5$ (in LB units) for left, middle and right columns, respectively. For each value of the friction the top row is a colour map indicating variations in concentration and the bottom row is the corresponding director field coloured by the orientation of nematic directors. $+1/2$ and $-1/2$ defects are denoted by red circles and blue triangles. (**c**) The RMS velocity is reduced by increasing friction. The inset illustrates the variation in log–log plot, showing the existence of different exponents. (**d**) The total number of topological defects initially increases, but drops sharply at $\gamma \sim 0.1$ and disappears at the dry limit.

(low) concentrations. Note the striking difference between the wet and dry limits: in the wet limit, concentration bands are initially formed perpendicular to the director, and the concentration and the nematic order are anti-correlated. In the dry limit, the bands form parallel to the director, and the concentration and nematic fields are correlated. The variation of the RMS velocity suggests the existence of different regimes (Fig. 2c, inset). From equation (8), the magnitude of the velocity can be estimated as $u_{RMS} \approx \zeta q/\gamma L_q$, where $L_q$ is the characteristic length scale of the nematic order variation. When the frictional damping is strong, $L_q$ is set by the distance between the walls, which is approximately proportional to the screening length $L_{sc} \sim \sqrt{\eta/\gamma\rho}$ and, therefore, the velocity is expected to vary as $u_{RMS} \sim \gamma^{-1/2}$ (Fig. 2c, inset: green line). At even higher frictions, there is not enough energy for hydrodynamic wall formation and the active flux-induced instabilities of dry matter dominate. Therefore, $L_q$ does not depend on the screening length and $u_{RMS} \sim \gamma^{-1}$ (Fig. 2c, inset: blue line).

### Flow-vortex lattices and ordered topological defects
We next focus on the dynamical effects of friction at a temperature corresponding to the (passive) isotropic–nematic transition (Fig. 3). In a passive nematic there would be no orientational order. However, order can be stabilized as a result of extensional flows generated by the active forcing for $\zeta\gamma > 0$, that is for an extensile, rod-like system ($\zeta > 0$, $\lambda > 0$) or a contractile, disc-like system ($\zeta < 0$, $\lambda < 0$)[42].

Considering first extensile rods, for the smallest value of friction considered active turbulence is established. Flow-vortices move around chaotically, decay and re-form (Fig. 3a and Supplementary Movies 1 and 2). As the friction coefficient is increased there is a striking change in behaviour. As is apparent in Fig. 3b, the velocity vortices form ordered, stationary arrays (Supplementary Movies 3 and 4). Figure 3c indicates how the vortices become narrower at higher friction.

The vortex lattice is interleaved with an ordered network of topological defects (Fig. 3e,f). The defect lattice is directly linked to the velocity pattern: $+1/2$ defects are generated between pairs of counter-rotating vortices due to the change in the sign of the vorticity[24], while $-1/2$ defects are created between two pairs of co-rotating vortices due to distinct flow-induced reorientation of nematic directors (Fig. 3g). The shear flow in the middle of a vortex turns a nematogen towards a 45° orientation relative to the vortex line, while extensional and compressional flows at the edges of a vortex induce vertical and horizontal reorientations, respectively.

Similar behaviour is seen for a system of contractile, disk-like particles. In the absence of friction, the activity-induced ordering results in the generation of turbulent-like flow patterns (Fig. 4a,b). As the friction coefficient is increased, vortices arrange to form a lattice (Fig. 4c–e). The long-range order of the vortex lattice is evident from measurements of the vorticity–vorticity correlation functions (Fig. 4f), which are defined as $C_{\omega\omega}(r) = \langle \boldsymbol{\omega}(r,t) \cdot \boldsymbol{\omega}(0,t) \rangle / \langle \boldsymbol{\omega}(0,t)^2 \rangle$, where $\boldsymbol{\omega}$ is the vorticity. A similar ordered array of vortices has recently been reported in experiments on dividing endothelial cells[19] where the vortex structure was associated with the division-induced flow field.

In the absence of friction, the vorticity length scale is set by the activity and the elastic constant[13,37]. However, for both extensile rod-like and contractile disk-like particles, on increasing friction, the vorticity length scale drops. When it becomes comparable to the screening length vortices are no longer disturbed by the flow and the vortex lattice is established (Fig. 5a).

The positional ordering of defects for extensile and contractile systems can be compared by calculating the structure factor $S_\mathbf{k} = \sum_{n=1}^{N_d} e^{2\pi i (\mathbf{k} \cdot \mathbf{x}_n)}$, where $\mathbf{x}_n$ denotes the position of the $n$th defect[43]. At low friction, positions of topological defects are uncorrelated (Figs 3d and 5b,c; low friction), while positional ordering is observed at high friction values (Figs 3e,f and 5b,c; high friction). The mechanism of the ordering is different in extensile and contractile systems as the primary mode of instability in contractile systems is a splay distortion of the director field, while in extensile systems, bend distortions dominate[44]. This manifests itself in different structure factors of defect ordering for extensile rod-like and contractile disk-like particles, which correspond to rectangular and centred rectangular Bravais crystal structures[43], respectively (Fig. 5b,c).

### Discussion
In addition to the vortex lattice surrounding epithelial cells, a similar positional ordering of defects has been observed in active polar systems in the form of aster arrays[45–47]. Moreover, the vorticity pattern we report at high friction resembles self-organized vortices of sperm cells on a surface[48] and periodic vortex arrays reported for a motility assay of microtubules with short-range attractions[7]. Recently, orientational, but not positional, ordering of topological defects has been observed in experiments on microtubule bundles driven by molecular motors[49], but the mechanism for this is not yet understood.

Friction is present in many active systems and can be tuned, providing mechanistic insights into pattern formation in active materials. For example, in the experiments reported in ref. 6, the microtubule bundles and kinesin molecular motors move on an oil–water interface. The active layer experiences friction due to the momentum transfer to the surrounding fluid and thus the friction can be varied by changing the viscosity difference between the upper and lower fluid layers[50]. In addition, experimental measurements of the flow fields of endothelial cells[19] could be extended by exploring the flow patterns on substrates with different coatings to alter the frictional properties. Indeed recent studies have demonstrated the important role of friction in altering the self-propulsion mechanism of a single cell in confined migration[51] and have shown that the cortical friction can stabilize actin patterns in epithelial tubes[52]. The latter study shows that in the absence of orientational order, the combined effects of friction, actin treadmilling and myosin contractility control the formation of the actin ring on the cell cortex[52].

The ability to control flow vorticity and defect structure will be important in the design and operation of biological and biomimetic materials. Scientists are just starting to try to create active machines that mimic nature on the microscopic scale[10] and active matter is being considered by those traditionally interested in device physics and novel materials[53]. The formation of the vortex lattice leads to the positional ordering of topological defects that could provide an early step towards the design and control of active materials producing well defined velocity fields. For example, one might envisage that such a vortex lattice has the potential to drive an array of microscopic gears.

### Methods
**Non-dimensionalization.** To characterize the relative importance of different stress contributions, we turn to the non-dimensional form of the momentum equation (equation (4)) introducing the characterisitic length and characterisitc velocity as $L$, $U$, respectively, we have $\mathbf{x}^* = \mathbf{x}L$, $\mathbf{u}^* = \mathbf{u}U$ and $p^* = p/(\rho U^2)$, where * denotes dimensionless variables. Equation (4) is then written in

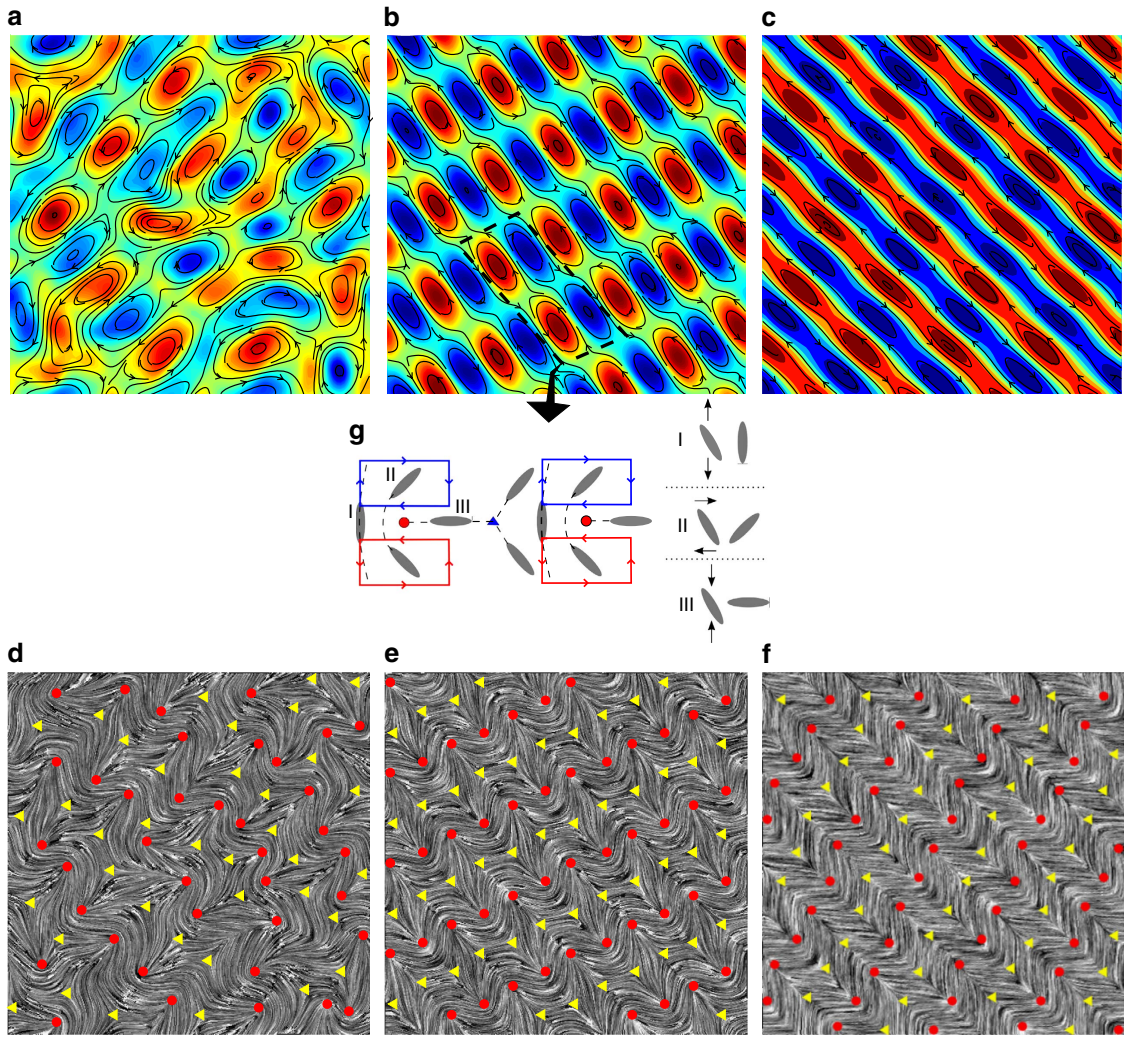

**Figure 3 | Emergence of a vortex lattice and defect ordering for extensile rod-like particles.** (**a–c**) Velocity field coloured by the magnitude of the vorticity. (**d–f**) Director fields visualized by Line Integral Convolution and superimposed by topological defects (with $+1/2$ and $-1/2$ defects denoted by red circles and yellow triangles). The hydrodynamic screening length is $L_{sc} = 15.30, 7.51, 5.10$, (lattice units) for the left, middle, and right columns, respectively. (**g**) Orientation of a defect is determined by its position relative to neighbouring vortices. A director responds differently to either extensional (I) or compressional (III) flow at the edges of vortices while it experiences a shear (II) inside a vortex.

non-dimensional form as

$$\partial_t^* \mathbf{u}^* + \mathbf{u}^* \cdot (\nabla^* \mathbf{u}^*) = -\nabla^* p^* + \nabla^* \cdot \left( \frac{1}{Re} \mathbf{\Pi}^{\text{viscous}*} + \frac{1}{ReEr} \mathbf{\Pi}^{\text{elastic}*} \right.$$
$$\left. + \frac{1}{Re} \frac{\zeta L}{\eta U} \mathbf{\Pi}^{\text{active}*} + \frac{1}{Re} \frac{\rho \gamma L^2}{\eta} \mathbf{\Pi}^{\text{friction}*} \right). \tag{11}$$

It shuold be noted that $1/Re$ where $Re = \rho UL/\eta$, is the Reynolds number, is the common prefactor for all the stress terms. For the parameters that we use in our simulations, both active stress ($\zeta L/\eta U$) and frictional stress ($\rho \gamma L^2/\eta$) are of the same order for large enough friction ($>O(0.01)$), and much larger than unity. The viscous stresses are on the order of $1/Re$ and the elastic terms are on the order of $1/(ReEr)$, where $Er = \eta UL/K$ is the Ericksen number. The Ericksen number characterizes the ratio of viscous to elastic forces and is on the order of $O(10^2)$ in our simulations. Thus both viscous and elastic stresses are not important for large values of friction coefficient.

**The high friction limit.** When the energy generated by the active particles is primarily dissipated as friction we may write $\gamma \mathbf{u} \approx -\zeta \nabla \cdot \mathbf{Q}$, which gives an expression for the velocity field in terms of distortions in the nematic director field, $\mathbf{u} \approx -(\zeta/\gamma) \nabla \cdot \mathbf{Q}$. Considering this as an equality and substituting into the equation (1) for the concentration evolution gives

$$\partial_t \phi = \Gamma_\phi a \nabla^2 \phi + \frac{\zeta \phi}{\gamma} \partial_i \partial_j Q_{ij} + \frac{\zeta}{\gamma} (\partial_j Q_{ij})(\partial_i \phi)$$
$$+ \Gamma_\phi \nabla^2 \left[ \frac{A}{2} \mathbf{Q}^2 - K_\phi \nabla^2 \phi + K_c \partial_i \partial_j Q_{ij} \right]. \tag{12}$$

In this equation, the terms in $\Gamma_\phi$ represent diffusive dynamics arising from free energy gradients. The second term on the right-hand side is the current of active particles due to self-generated flow. This is usually introduced as a phenomenological, curvature-driven current in models of dry active nematics[3,21,23]. The third term is a non-linear term arising again from advection of concentration but rarely used in models of dry active systems.

Similarly, substituting for the velocity field in the equation (2) for the nematic order parameter we obtain

$$\partial_t Q_{ij} = \Gamma_Q \left[ \left( -A\phi - CQ_{pq}Q_{qp} \right) Q_{ij} + K_Q \partial_\gamma^2 Q_{ij} + K_c \left( \frac{1}{2} \delta_{ij} \partial_k^2 - \partial_i \partial_j \right) \phi \right]$$
$$+ \frac{\zeta}{\gamma} \left[ \partial_m Q_{km} \partial_k Q_{ij} \right] - \frac{\zeta}{2\gamma} \left[ \lambda \partial_m (\partial_i Q_{km} + \partial_k Q_{im}) + \partial_m (\partial_k Q_{im} - \partial_i Q_{km}) \right] \left[ Q_{kj} + \frac{\delta_{kj}}{3} \right]$$
$$- \frac{\zeta}{2\gamma} \left[ Q_{ik} + \frac{\delta_{ik}}{3} \right] \left[ \lambda \partial_m (\partial_k Q_{jm} + \partial_j Q_{km}) - \partial_m (\partial_j Q_{km} - \partial_k Q_{jm}) \right]$$
$$+ \frac{2\lambda \zeta}{\gamma} \left[ Q_{ij} + \frac{\delta_{ij}}{3} \right] Q_{kl} \partial_k \partial_m Q_{ml}. \tag{13}$$

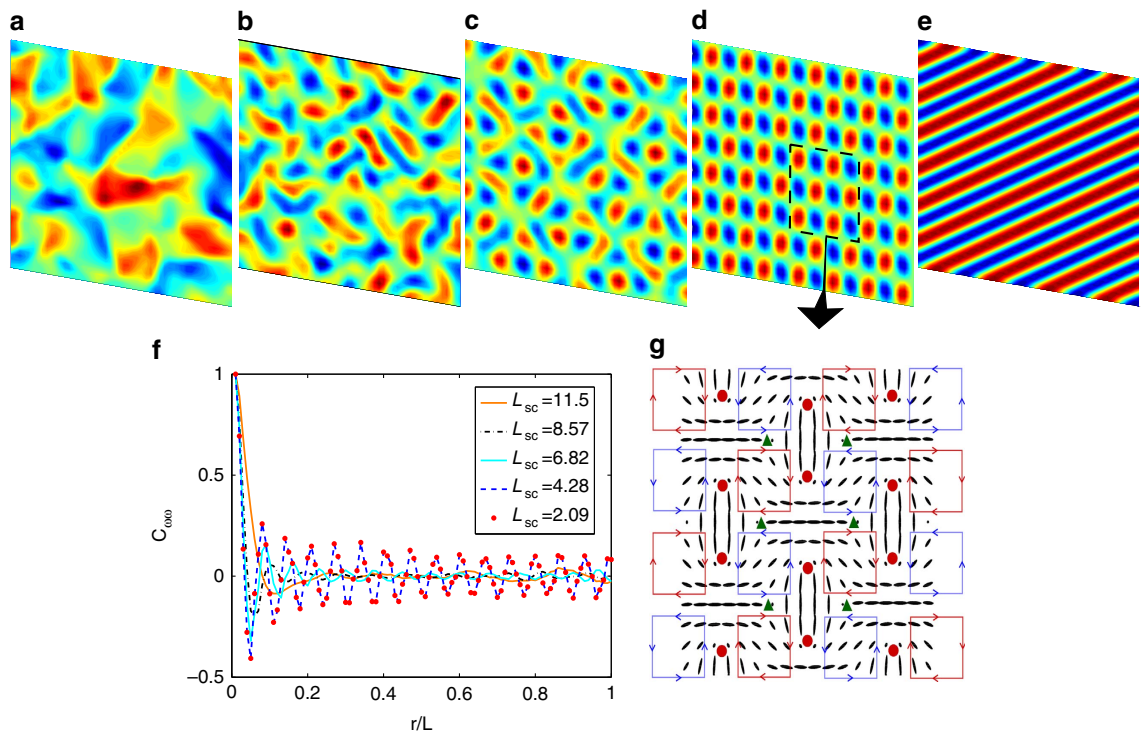

**Figure 4 | Emergence of a vortex lattice and defect ordering for contractile disk-like particles.** (**a**–**e**) Velocity field coloured by the magnitude of the vorticity. The hydrodynamic screening length is $L_{sc} = 11.5, 8.57, 6.82, 4.28, 2.09$ (lattice units) for (**a**–**e**) respectively. (**f**) Vorticity–vorticity correlations function $C_{\omega\omega}$ demonstrates the transition from active turbulence state to the vortex lattice configuration. (**g**) The stable structure of nematic directors and topological defects in the vortex lattice. Solid red and blue lines illustrate the clockwise and counterclockwise vortices. Topological defects with charges $+1/2$, $-1/2$, are shown by red circles and green triangles, respectively.

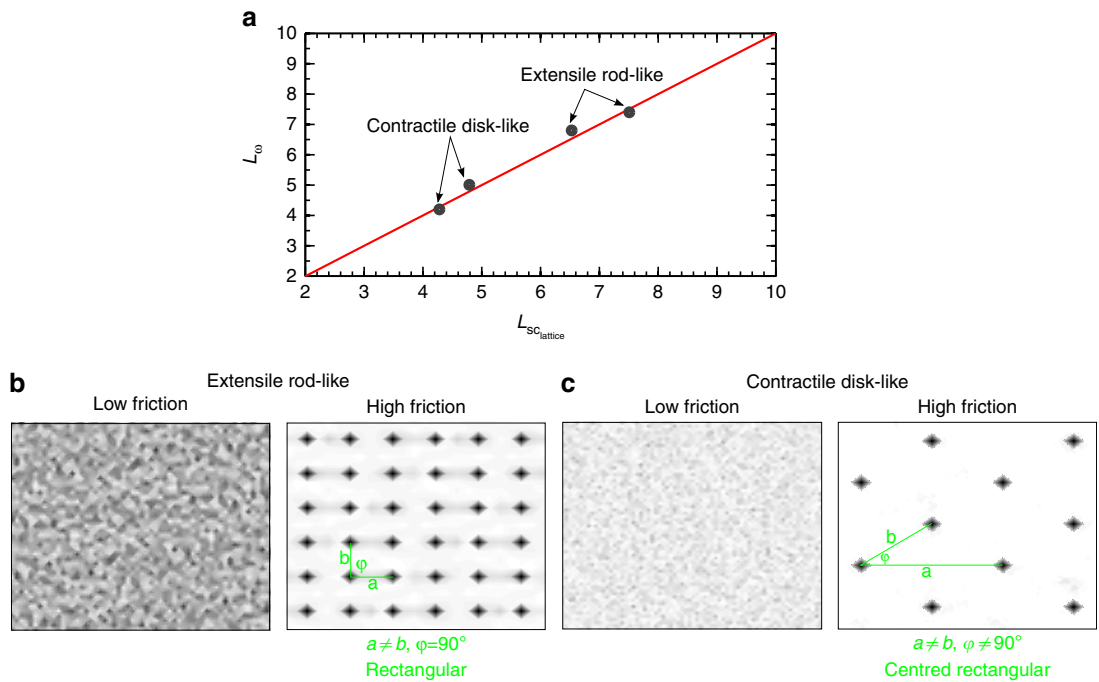

**Figure 5 | Properties of the vortex lattice.** (**a**) Dots indicate the characteristic vorticity length scale and the screening length at which the vortex lattice emerges for extensile and contractile systems, each for two different values of the activity. The red line indicates $L_\omega = L_{sc}$. (**b**,**c**) Colormaps of the structure factor for defects showing the emergence of positional ordering of defects at high friction for (**b**) extensile rod-like and (**c**) contractile disk-like particles.

Here, the terms written on the first line are generally used to describe the dynamics of dry active nematics. Terms on the other lines, that are proportional to $\zeta/\gamma$, are generated by the presence of active flow and flow gradients, just like the curvature-driven term that appeared in equation (6) in the main text. These contributions are not usually considered in dry active nematics [3,21], because they represent non-linear contributions which are assumed to be sub-dominant in defining the dynamics. We now follow the approach of Bertin et al. [21] to show explicitly that such flow coupling terms can arise in general descriptions of dry systems.

**Microscopic derivation of the flow-driven terms.** We briefly review the microscopic approach used by Bertin et al. [21] to derive the equations of motion of dry active nematics. (For consistency, we use the relative concentration $\phi$ rather than the number density used in ref. 21.) Bertin et al. consider $N$ particles which are characterized by position $\mathbf{x}$ and orientation $\theta$. The alignment interactions are similar to those in a Vicsek model [20]; however there is no net motion along the orientation vector of each particle, thus imposing nematic symmetry. Considering the motion of the particles, taking into account their angular diffusion and inter-particle collisions, Bertin et al. showed that the evolution of the Fourier transform components of the single particle probability distribution $\hat{f}_k(\mathbf{x},\theta,t)$ follow the master equations

$$\partial_t \hat{f}_k(\mathbf{x},t) = \frac{1}{2}\tilde{\nabla}\tilde{\nabla}^* \hat{f}_k(\mathbf{x},t) + \frac{1}{4}\left(\tilde{\nabla}^{*2}\hat{f}_{k+1} + \tilde{\nabla}^2 \hat{f}_{k-1}\right) + [\hat{P}_k - 1]\hat{f}_k(\mathbf{x},t) \\ + \frac{1}{\pi}\sum_q J(\hat{P}_k,k,q)\hat{f}_q(\mathbf{x},t)\hat{f}_{k-q}(\mathbf{x},t), \quad (14)$$

where $\hat{P}_k$ is the Fourier transform of the noise distribution for angular displacements and $J(\hat{P}_k,k,q) = 4\hat{P}_k \frac{1+w\sqrt{2}(2q-k)(-q)^q \sin(k\pi/2)}{1-4(2q-k)^2} - \frac{4}{1-16q^2}$. Also, $\tilde{\nabla} = \partial_x + i\partial_y$ and $\tilde{\nabla}^*$ is the complex conjugate of $\tilde{\nabla}$.

To close this infinite hierarchy of equations, a small parameter $\epsilon$ is introduced that characterizes the distance to the isotropic–nematic transition. Assuming that $\hat{f}_k \sim |\epsilon|^k, \partial_t \sim \epsilon^2, \partial_x \sim \epsilon$, the first two modes ($k=0,1$) are given by

$$\partial_t \hat{f}_0(\mathbf{x},t) = \frac{1}{2}\tilde{\nabla}\tilde{\nabla}^* \hat{f}_0(\mathbf{x},t) + \frac{1}{4}\left[\tilde{\nabla}^{*2}\hat{f}_1 + \tilde{\nabla}^2 \hat{f}_1^*\right], \quad (15)$$

$$\partial_t \hat{f}_1(\mathbf{x},t) = \frac{1}{2}\tilde{\nabla}\tilde{\nabla}^* \hat{f}_1(\mathbf{x},t) + \frac{1}{4}\tilde{\nabla}\tilde{\nabla}\hat{f}_0 \\ + A'\hat{f}_1 + C'\hat{f}_1^*\hat{f}_1^2 + D'\tilde{\nabla}^*\tilde{\nabla}^*\hat{f}_1^2 + E'\hat{f}_1\tilde{\nabla}\tilde{\nabla}\hat{f}_1, \quad (16)$$

where the equations are truncated $O(\epsilon^4)$ and the $k=2$ mode is used to close the equations. Coefficients $A', C', D'$ and $E'$ represent material parameters, that may also depend on $\rho$. Identifying the zeroth mode as the concentration ($\hat{f}_0 = \phi$) and first mode as the concentration weighted nematic field ($\hat{f}_1 = 2(\phi Q_{xx} + i\phi Q_{xy})$), Bertin et al. [21] obtained the following equations $O(\epsilon^3)$:

$$\partial_t \phi = \frac{1}{2}\nabla^2 \phi + \partial_i \partial_j (\phi Q_{ij}), \quad (17)$$

$$\partial_t (\phi Q_{ij}) = A'(\phi Q_{ij}) + \frac{C'}{2}\left[(\phi Q_{pq})(\phi Q_{qp})\right](\phi Q_{ij}) \\ + \frac{1}{2}\nabla^2 (\phi Q_{ij}) - \frac{1}{4}\left(\frac{\delta_{ij}}{2}\partial_k^2 - \partial_i \partial_j\right)\phi. \quad (18)$$

These are the familiar equations describing dry active nematics, see eg equations (6,7) in the main text or the first lines of equations (12,13) (A minor difference may be attributed to the approaches adopted: as is usual in kinetic descriptions the equations describe the evolution of $\phi Q$ rather than $Q$, while continuum approaches are generally not written in terms of concentration weighted fields.)

The two additional terms that appear $O(\epsilon^4)$ in the analysis are $D'\tilde{\nabla}^*\tilde{\nabla}^*\hat{f}_1^2 + E'\hat{f}_1\tilde{\nabla}\tilde{\nabla}\hat{f}_1$, the last two terms of equation (16). These terms, that are not used in Bertin et al.[21] have the role of advecting and co-rotating $\mathbf{Q}$ with the flow field. They give rise to all the non-linear terms in the last three lines of equation (13) except those proportional to the alignment parameter $\lambda$. Thus we establish that the flow-driven terms in equation (13) can be present in the kinetic description of dry active systems, but appear only at higher orders.

It is interesting to note that kinetic theory treats the case $\lambda=0$, that is, the role of the symmetric part of the velocity gradient tensor is not taken into consideration. Further physics will need to be incorporated in the kinetic approach to capture the flow aligning behaviour of active nematics (terms in the equations of motion that depend on $\lambda$). Similarly modifying the interactions between the particles may also yield additional coupling and the higher order diffusive terms that appear on the second line of equation (12).

**Numerics.** The governing equations of active nematohydrodynamics (equations (1–4)) describe a coupled system of partial differential equations. We employ a hybrid lattice Boltzmann technique to solve them. In this method, the equations describing the evolution of density and momentum (equations (3,4)) are solved in tandem using a discretized version of the Boltzmann equation where the first and second moments of the particle distribution function give the density and momentum respectively. The Bhatnagar–Gross–Krook approximation with a single relaxation time is used in the collision operator. Equations (1,2) which respectively describe the evolution of the concentration and orientational order parameter are solved using the method of lines. All spatial differentials are discretized using the second order central difference scheme and time integration is performed using an Euler method. The time step for the method of lines is chosen as 1/10th of that for the lattice Boltzmann updates. The flow field used to evolve the order parameters is updated after every lattice Boltzmann time step while the active and passive stress are determined using the updated order parameter fields in every lattice Boltzmann step, hence ensuring the coupling between the equations in the algorithm. As usual in the lattice Boltzmann technique, discrete space and time steps are chosen as unity. Details can be found in refs 24,31. Simulations are performed on a $100 \times 100$ lattice. Unless otherwise stated, the parameters used are $\Gamma_\phi = 1.0$, $\Gamma_Q = 0.1$, $a = 0$, $A = -1.0$, $C = 0.6$, $K_\phi = 0.1$, $K_Q = 0.01$, $K_c = 0.0$, and $\eta = 0.05$, in lattice units. We use $\zeta = 0.008$, $\lambda = 0.7$ for extensile, rod-like particles and $\zeta = -0.02$, $\lambda = -0.3$ for contractile, disk-like active systems.

It is noteworthy that material parameters in active systems are not known experimentally, and the research in this direction is in progress. In our simulations, we have used the usual lattice Boltzmann units[31], and therefore an experimentalist may suitably dimensionalize the system of interest by choosing a length, time and mass scale[24,54].

### Acknowledgements

We acknowledge funding from the ERC Advanced Grant MiCE. We thank Tyler Shendruk and Ramin Golestanian for helpful conversations.

### Author contributions

S.P.T., A.D. and J.M.Y. conceived the research and J.M.Y. and S.P.T. supervised the project. A.D. carried out the simulations. All authors contributed to performing analytical calculations, interpreting the results and writing the paper.

### Additional information

**Supplementary Information** accompanies this paper at http://www.nature.com/naturecommunications

**Competing financial interests:** The authors declare no competing financial interests.

**Reprints and permission** information is available online at http://npg.nature.com/reprintsandpermissions/

**How to cite this article:** Doostmohammadi, A. *et al.* Stabilization of active matter by flow-vortex lattices and defect ordering. 7:10557 doi: 10.1038/ncomms10557 (2016).

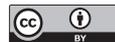